\documentclass[prl,superscriptaddress,amsmath,amssymb,twocolumn]{revtex4}
\usepackage{bm}
\usepackage{graphicx}

\begin{document}
\newcommand{\tr}{\mathop{\mathrm{tr}}\nolimits}
\newcommand{\adj}{\mathop{\mathrm{adj}}}
\newcommand{\diag}{\mathop{\mathrm{diag}}}
\renewcommand{\Re}{\mathop{\mathrm{Re}}}
\renewcommand{\Im}{\mathop{\mathrm{Im}}}
\newtheorem{lemma}{Proposition}

\preprint{}
\title{Bifurcation Phenomenon in a Spin Relaxation}
\author{Gen Kimura}
\email{gen@hep.phys.waseda.ac.jp}
\author{Kazuya Yuasa}
\email{yuasa@hep.phys.waseda.ac.jp}
\affiliation{Department of Physics, Waseda University, Tokyo 169--8555, Japan}
\author{Kentaro Imafuku}
\email{imafuku@volterra.mat.uniroma2.it}
\affiliation{Centro Vito Volterra, Universit\`a di Roma Tor Vergata, 00133 Rome, Italy}
\date[]{September 3, 2001}
\begin{abstract}
Spin relaxation in a strong-coupling regime (with respect to the spin system) is investigated in detail based on the spin-boson model in a stochastic limit.
We find a bifurcation phenomenon in temperature dependence of relaxation constants, which is never observed in the weak-coupling regime.
We also discuss inequalities among the relaxation constants in our model and show the well-known relation $2\mathnormal{\Gamma}_T\ge\mathnormal{\Gamma}_L$, for example, for a wider parameter region than before.
\end{abstract}
\maketitle

Nowadays, we are gaining technologies for controlling quantum states, and there are increasing necessities for better understandings on dissipation or relaxation in nanoscale systems.
Many authors have been working on this kind of issue, and we already have some knowledge about it, especially in the field of quantum optics, where the weak-coupling limit, e.g., the van Hove limit~\cite{ref:VanHove,ref:Spohn}, is a very good approximation.
There are, however, many other situations we have to handle.

In this Letter, we focus our attention on a dissipative dynamics derived through a different scaling limit from that of van Hove's;
the limit $\lambda\to0$ with the scaled time $\tau=\lambda^2t$ fixed is applied to a total Hamiltonian of the form~\cite{ref:Spohn,ref:Palmer}
\begin{equation}
H_\mathrm{tot}=\lambda^2H_S+\lambda V+H_B,
\label{eqn:RapidDecayHamiltonian}
\end{equation}
where $H_S$ is the Hamiltonian of a dissipative system of interest, $H_B$ that of an environment with infinite degrees of freedom, and $\lambda V$ the interaction Hamiltonian.
In contrast to the van Hove limit~\cite{ref:VanHove,ref:Spohn}, which is applied to a Hamiltonian of the form $H_\mathrm{tot}=H_S+\lambda V+H_B$, the limit for the Hamiltonian~(\ref{eqn:RapidDecayHamiltonian}) is regarded as a \textit{strong-coupling limit} in the sense that the interaction $\lambda V$ is much larger than the energy scale of the system $\lambda^2H_S$.

The systems with the scaling~(\ref{eqn:RapidDecayHamiltonian}) have been discussed by several people~\cite{ref:Spohn,ref:Palmer}, and their importance is recognized already.
The present authors have recently constructed~\cite{ref:SLAForRapidDecay} a framework of the stochastic limit for the scaling~(\ref{eqn:RapidDecayHamiltonian}), modifying the original stochastic limit which is proposed by Accardi~\textit{et~al.}~\cite{ref:SLACommunMathPhys,ref:SLAinSpinBoson} on the bassis of the van Hove limit.
Applying the framework for the scaling~(\ref{eqn:RapidDecayHamiltonian}) to the spin-boson model~\cite{ref:SpinBoson}, which is simple but is fundamental and has many applications, we here investigate the spin relaxation under the scaling~(\ref{eqn:RapidDecayHamiltonian}) in detail and report interesting features in the relaxation dynamics, which are never found in the weak-coupling regime.

We consider the spin-boson model~\cite{ref:SpinBoson}
\begin{subequations}
\label{eqn:SpinBosonModel}
\begin{equation}
H_S=\frac{\varepsilon}{2}\sigma_z+\frac{\mathnormal{\Delta}}{2}\sigma_x,
\quad
H_B=\int_0^\infty d\omega\,\hbar\omega a_\omega^\dagger a_\omega,
\end{equation}
\begin{equation}
V=i\hbar\sigma_z\int_0^\infty d\omega\,
\Bigl(g(\omega)a_\omega-g^*(\omega)a^\dagger_\omega\Bigr)
\end{equation}
\end{subequations}
under the scaling~(\ref{eqn:RapidDecayHamiltonian}).
Here $\sigma_x$, $\sigma_z$ are the Pauli matrices, and the spin system $H_S$ has two eigenstates
\begin{equation}
H_S|\pm\rangle=\pm\frac{1}{2}\hbar\mathnormal{\Omega}_0|\pm\rangle,\quad
\hbar\mathnormal{\Omega}_0=\sqrt{\varepsilon^2+\mathnormal{\Delta}^2}.
\end{equation}
The environment $H_B$ is the boson system, where $a_\omega$ ($a_\omega^\dagger$) is the annihilation (creation) operator of the boson of mode $\omega$.
Two systems interact with each other through the linear coupling interaction $V$.
In the following analysis, the form factor $g(\omega)$ is assumed to be such that $2\pi|g(\omega)|^2\sim\eta\omega$ for $\omega\sim0$ (Ohmic model).
We assume further that, before time $t=0$, the spin and boson systems are uncorrelated and the boson system is in the thermal equilibrium state at temperature $T$, i.e., the initial state of the total system is given by
\begin{equation}
\rho_\mathrm{tot}(0)=\rho_S\otimes\rho_B,\quad
\rho_B=\frac{1}{Z}e^{-H_B/k_BT},
\end{equation}
where $k_B$ is the Boltzmann constant and $Z$ the normalization constant.

Under these assumptions, it is possible to derive a master equation for the density operator of the spin system, $\rho_S(\tau)=\tr_B\rho_\mathrm{tot}(\tau)$.
In the stochastic limit for the scaling~(\ref{eqn:RapidDecayHamiltonian}), the master equation reads~\cite{ref:SLAForRapidDecay}
\begin{equation}
\frac{d}{d\tau}\rho_S(\tau)
=-\frac{i}{\hbar}[H_S,\rho_S(\tau)]
-\frac{\gamma^\theta}{4}\bm{[}\sigma_z,[\sigma_z,\rho_S(\tau)]\bm{]},
\label{eqn:MasterEq}
\end{equation}
where
\begin{equation}
\gamma^\theta=\frac{2\eta k_BT}{\hbar}.
\end{equation}
The equivalent \textit{Bloch equation} for the spin system operators $D_\pm=|\pm\rangle\langle\mp|$, $D_0=|+\rangle\langle+|-|-\rangle\langle-|$ is given, through the relations $\tr_S[\rho_S(\tau)D_\pm]=\tr_S[\rho_SD_\pm(\tau)]$, $\tr_S[\rho_S(\tau)D_0]=\tr_S[\rho_SD_0(\tau)]$, by
\begin{subequations}
\label{eqn:BlochEq}
\begin{equation}
\frac{d}{d\tau}\bm{D}(\tau)=-A\bm{D}(\tau),
\end{equation}
where
\begin{widetext}
\begin{equation}
\bm{D}(\tau)=\left(
\begin{array}{cc}
\medskip
D_+(\tau)\\
\medskip
D_-(\tau)\\
D_0(\tau)
\end{array}
\right),\quad
A=\left(
\begin{array}{ccc}
\medskip
(\tilde{\mathnormal{\Delta}}^2+2\tilde{\varepsilon}^2)\gamma^\theta/2
-i\mathnormal{\Omega}_0&
-\tilde{\mathnormal{\Delta}}^2\gamma^\theta/2&
-\tilde{\varepsilon}\tilde{\mathnormal{\Delta}}\gamma^\theta/2\\
\medskip
-\tilde{\mathnormal{\Delta}}^2\gamma^\theta/2&
(\tilde{\mathnormal{\Delta}}^2+2\tilde{\varepsilon}^2)\gamma^\theta/2
+i\mathnormal{\Omega}_0&
-\tilde{\varepsilon}\tilde{\mathnormal{\Delta}}\gamma^\theta/2\\
-\tilde{\varepsilon}\tilde{\mathnormal{\Delta}}\gamma^\theta&
-\tilde{\varepsilon}\tilde{\mathnormal{\Delta}}\gamma^\theta&
\tilde{\mathnormal{\Delta}}^2\gamma^\theta
\end{array}
\right),
\end{equation}
\end{widetext}
\end{subequations}
and $\tilde{\varepsilon}=\varepsilon/\hbar\mathnormal{\Omega}_0$, $\tilde{\mathnormal{\Delta}}=\mathnormal{\Delta}/\hbar\mathnormal{\Omega}_0$.
Based on the master equation~(\ref{eqn:MasterEq}) or the Bloch equation~(\ref{eqn:BlochEq}), we investigate a spin relaxation in the scaling~(\ref{eqn:RapidDecayHamiltonian}).

\begin{figure}[b]
\begin{tabular}{c@{\quad}c}
\includegraphics[height=0.105\textwidth]{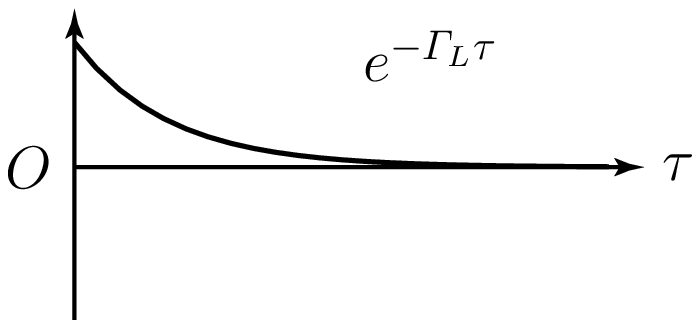}&
\includegraphics[height=0.105\textwidth]{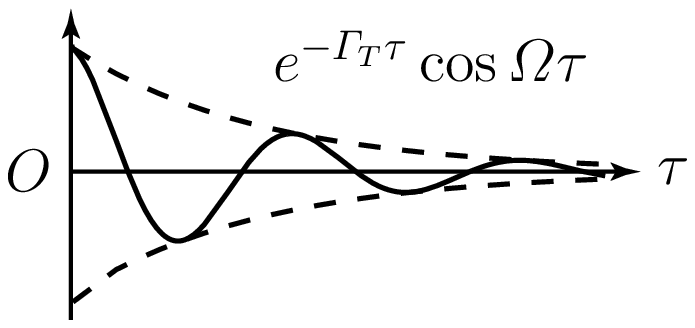}\\
(a)&(b)
\end{tabular}
\caption{(a) Longitudinal and (b) transverse relaxations}
\label{fig:LTImage}
\end{figure}

The dynamics of the spin system is characterized by the eigenvalues of the matrix $A$ in the Bloch equation~(\ref{eqn:BlochEq}).
There are three eigenvalues, which are analytically obtained by Cardano's formula for the solutions to third order polynomial equation.
In some cases, two of them are complex conjugate to each other, and the remaining one is real.
The real eigenvalue gives an exponential decay, and the complex ones damped oscillations (Fig.~\ref{fig:LTImage}).
We call the former ``longitudinal relaxation'' and the latter ``transverse relaxation.''
The damping rates $\mathnormal{\Gamma}_L$ for the longitudinal relaxation and $\mathnormal{\Gamma}_T$ for the transverse one are defined, respectively, by
\begin{equation}
\left\{
\begin{array}{l}
\medskip
\mathnormal{\Gamma}_L\equiv(\text{real eigenvalue}),\\
\mathnormal{\Gamma}_T
\equiv(\text{real part of the complex eigenvalues}).
\end{array}
\right.
\end{equation}

Our definitions of the ``longitudinal'' and ``transverse'' relaxations seem formal, and the physical meaning of them is not so clear as is in the weak-coupling regime.
Indeed, in the latter case, the matrix $A$ in the corresponding Bloch equation is always diagonal $A=\diag(\mathnormal{\Gamma}_T-i\mathnormal{\Omega}_R,\mathnormal{\Gamma}_T+i\mathnormal{\Omega}_R,\mathnormal{\Gamma}_L)$ with $\mathnormal{\Gamma}_L=\tilde{\mathnormal{\Delta}}^2\gamma^\theta$, $\mathnormal{\Gamma}_T=\tilde{\mathnormal{\Delta}}^2\gamma^\theta/2$ [$\gamma^\theta=2\pi|g(\mathnormal{\Omega}_0)|^2\coth(\hbar\mathnormal{\Omega}_0/2k_BT)$ and $\mathnormal{\Omega}_R$ is a (renormalized) frequency]~\cite{ref:SLAinSpinBoson}, and the longitudinal relaxation is nothing but the energy relaxation.
In the Bloch equation~(\ref{eqn:BlochEq}), $\mathnormal{\Gamma}_L$ stands just for a damping rate of an observable (not necessarily energy) which decays exponentially without oscillation, and it depends on temperature (see below).
Furthermore, there are cases where all three eigenvalues are real giving three ``longitudinal''-relaxation constants $\mathnormal{\Gamma}_L^{(i)}$ ($i=1,2,3$).
It might be inappropriate, therefore, to introduce such notion as ``longitudinal'' or ``transverse'' in this regime.
One should notice, however, that a quite nontrivial phenomenon, i.e., bifurcation in the relaxation dynamics, can be observed in terms of $\mathnormal{\Gamma}_L$ and $\mathnormal{\Gamma}_T$ as is seen in the following.
We are interested, not in the matter of definition, but in the fact that the phenomenon peculiar to nonlinear (and/or nonequilibrium) systems can be seen in this system.

Now, let us investigate the relaxation constants $\mathnormal{\Gamma}_L$ and $\mathnormal{\Gamma}_T$ in detail.
First of all, it should be mentioned that positivity of the relaxation constants is guaranteed since the master equation~(\ref{eqn:MasterEq}) is of the Lindblad form~\cite{ref:LindbladCommunMathPhys1976,ref:GoriniKossakowskiSudarshan}.
For an investigation of the matrix $A$ in the Bloch equation~(\ref{eqn:BlochEq}), the following proposition will be useful:
\begin{lemma}
\label{Proposition1}
\upshape Let
\begin{equation}
M_{3,R}=\{A:3\times3\ \text{matrix}|\tr A,\,\det A,\,\tr\adj A\in\mathbb{R}\},
\label{eqn:M3R}
\end{equation}
where $\adj A$ is the cofactor matrix of $A$ (adjugate of $A$) and $\mathbb{R}$ is the set of real numbers.
The necessary and sufficient condition for the real parts of the eigenvalues $\lambda_i$ ($i=1,2,3$) of $A\in M_{3,R}$ to be positive semi-definite
\begin{equation}
\Re\lambda_i\ge0\quad(i=1,2,3)
\end{equation}
is that
\begin{equation}
\tr A\ge0,\ %
\det A\ge0,\ %
\tr\adj A\ge0,\ \text{and}\ %
f(\tr A)\ge0,
\label{eqn:i}
\end{equation}
where $f(\lambda)=\det(\lambda-A)$.
\end{lemma}
(Its proof is quite elementary and straightforward~\cite{ref:GenInequalities}.)
The matrix $A$ in the Bloch equation~(\ref{eqn:BlochEq}) belongs to $M_{3,R}$ and satisfies the condition~(\ref{eqn:i}).
The real parts of the eigenvalues of $A$, i.e., the relaxation constants, are therefore positive semi-definite by Prop.~\ref{Proposition1}~\cite{note:EquilibriumState}.

\begin{figure}[t]
\includegraphics[width=0.48\textwidth]{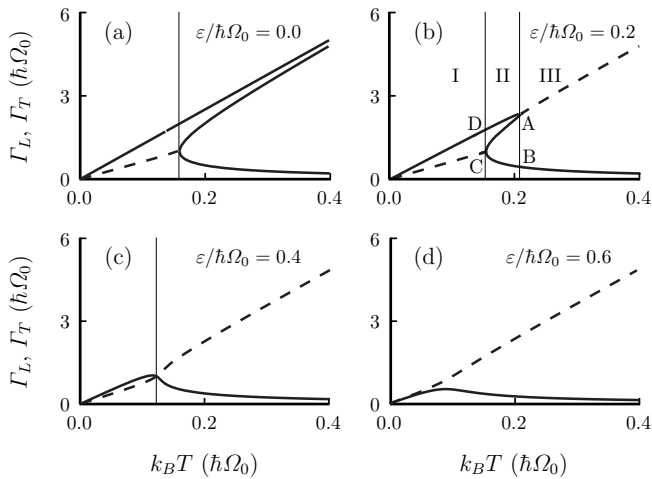}
\caption{Temperature dependences of the longitudinal-relaxation constant $\mathnormal{\Gamma}_L$ (solid lines) and the transverse one $\mathnormal{\Gamma}_T$ (dashed lines) for (a) $\varepsilon/\hbar\mathnormal{\Omega}_0=0.0$, (b) $0.2$, (c) $0.4$, and (d) $0.6$.}
\label{fig:RelaxationConstants}
\end{figure}
Next, let us discuss the parameter dependence of the relaxation constants $\mathnormal{\Gamma}_L$ and $\mathnormal{\Gamma}_T$.
Since we have derived the master equation~(\ref{eqn:MasterEq}) or the Bloch equation~(\ref{eqn:BlochEq}) from the microscopic point of view, i.e., from the basic Hamiltonian~(\ref{eqn:RapidDecayHamiltonian}) with~(\ref{eqn:SpinBosonModel}), the relaxation constants $\mathnormal{\Gamma}_L$ and $\mathnormal{\Gamma}_T$ are given in terms of the parameters in the Hamiltonian, $\varepsilon$, $\mathnormal{\Delta}$, $\eta$, and the temperature $T$.
Furthermore, since the relaxation constants are obtained analytically, their dependence on the parameters is clear.
In Fig.~\ref{fig:RelaxationConstants}, temperature dependences of the relaxation constants $\mathnormal{\Gamma}_L$ and $\mathnormal{\Gamma}_T$ are presented for various values of $\varepsilon/\hbar\mathnormal{\Omega}_0$.
The first point to notice is that they exhibit quite different behaviors depending on the parameter $\varepsilon/\hbar\mathnormal{\Omega}_0$.
This is not the case in the weak-coupling regime, where the relation $2\mathnormal{\Gamma}_T=\mathnormal{\Gamma}_L$ always holds.
In particular, there are temperature regions where \textit{no transverse-relaxation constant exists} [Figs.~\ref{fig:RelaxationConstants}(a) and~(b)].

Furthermore, observe an interesting and nontrivial phenomenon there.
See Fig.~\ref{fig:RelaxationConstants}(b), for example.
A transverse-relaxation constant bifurcates into two longitudinal ones at some temperature (at C from lower temperature and at A from higher temperature).
Or conversely, two longitudinal ones merge into one transverse one (at C from higher temperature and at A from lower temperature).

\begin{figure}[t]
\begin{tabular}{c@{\quad}c@{\quad}c}
\includegraphics[height=0.130\textwidth]{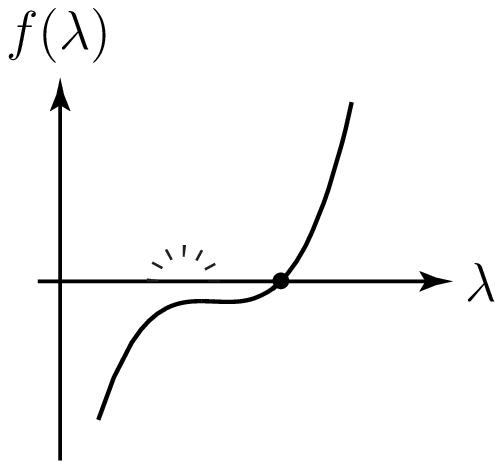}&
\includegraphics[height=0.130\textwidth]{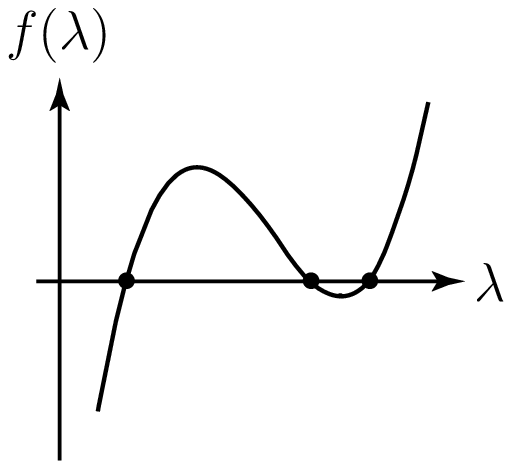}&
\includegraphics[height=0.130\textwidth]{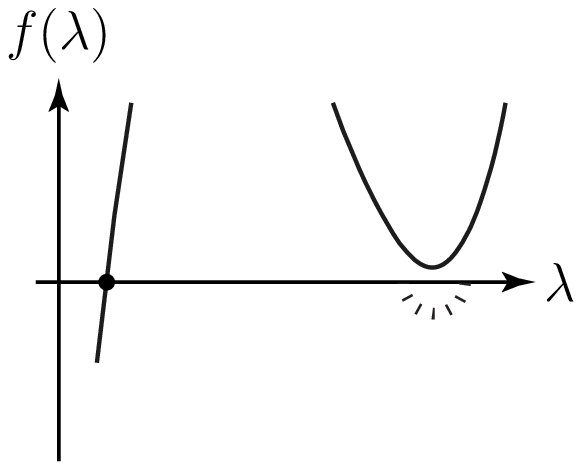}\\
(I)&(II)&(III)
\end{tabular}
\caption{Characteristic functions $f(\lambda)=\det(\lambda-A)$ in the temperature regions~I--III in Fig.~\ref{fig:RelaxationConstants}(b).}
\label{fig:Mechanism}
\end{figure}

The mechanism is understood by tracing the zero points of the characteristic function $f(\lambda)=\det(\lambda-A)$, i.e., the real eigenvalues of the matrix $A$.
The forms of $f(\lambda)$ in the temperature regions~I--III in Fig.~\ref{fig:RelaxationConstants}(b) are sketched in Fig.~\ref{fig:Mechanism}\@.
The real eigenvalue in region~I disappears in region~III and changes into two complex eigenvalues when the temperature becomes higher.
On the other hand, a new zero point comes out in region~II, i.e., two complex eigenvalues in region~I change into a real one.
These are typical bifurcation mechanisms, which are often seen in nonlinear (and/or nonequilibrium) systems~\cite{ref:PrigogineDissipativeStructure}.
This is a bifurcation phenomenon seen in a spin relaxation process.

\begin{figure}[t]
\includegraphics[width=0.350\textwidth]{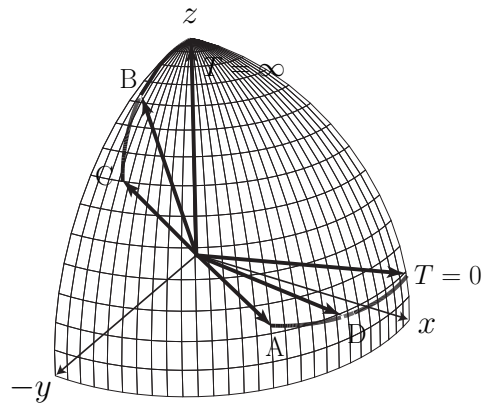}
\caption{Temperature dependence of the longitudinal direction $\bm{l}$ corresponding to Fig.~\ref{fig:RelaxationConstants}(b). In region~II, only two directions which connect continuously with that in region~I or~III are displayed.}
\label{fig:DiscontinuousBehavior}
\end{figure}

An experimental implication of this phenomenon is as follows.
Let us focus on a spin component
\begin{equation}
\sigma_L=\bm{l}\cdot\bm{\sigma}
=l_x\sigma_x+l_y\sigma_y+l_z\sigma_z
\end{equation}
which decays exponentially without oscillation: $\sigma_L(\tau)=\sigma_Le^{-\mathnormal{\Gamma}_L\tau}$.
The vector $\bm{l}$ is normalized to unity, and we call it ``longitudinal direction.''
Temperature dependence of the longitudinal direction $\bm{l}$ is shown in Fig.~\ref{fig:DiscontinuousBehavior}, corresponding to those of the relaxation constants in Fig.~\ref{fig:RelaxationConstants}(b).
In the limit $T\to0$, $\bm{l}$ is in the ``direction of the Hamiltonian'' ($\sigma_L\propto H_S$), and in the limit $T\to\infty$, it is in the ``direction of the interaction'' ($\sigma_L\propto V$).
If one traces $\bm{l}$ from $T=0$ to higher temperature, it disappears at some critical temperature [A in Fig.~\ref{fig:DiscontinuousBehavior} or~\ref{fig:RelaxationConstants}(b)] and is found in a different direction B\@.
If one traces it from higher temperature in reverse, it jumps at a different temperature from C to D\@.
The bifurcation phenomenon will be observed experimentally as a discontinuous temperature dependence of the longitudinal direction.

\begin{figure}[t]
\includegraphics[width=0.48\textwidth]{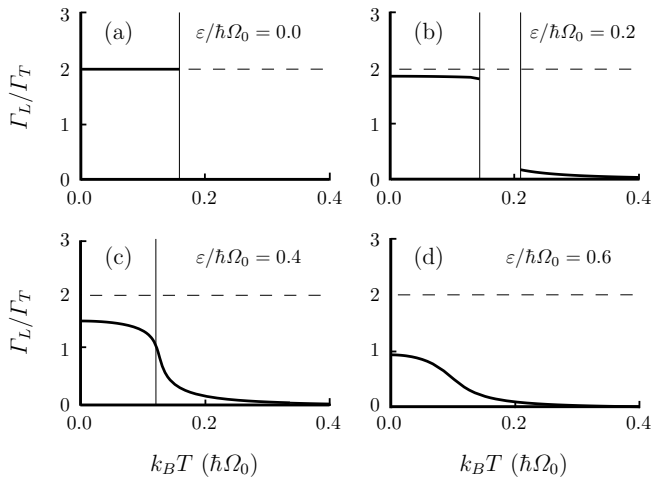}
\caption{Temperature dependences of $\mathnormal{\Gamma}_L/\mathnormal{\Gamma}_T$, corresponding to Fig.~\ref{fig:RelaxationConstants}. Note that there are regions where no transverse relaxation constant exists.}
\label{fig:RatefortheLandTrelaxationConsts}
\end{figure}

Let us finally discuss the ratio of the longitudinal-relaxation constant $\mathnormal{\Gamma}_L$ to the transverse one $\mathnormal{\Gamma}_T$.
Temperature dependences of it are plotted in Fig.~\ref{fig:RatefortheLandTrelaxationConsts}, corresponding to Fig.~\ref{fig:RelaxationConstants}\@.
They imply that the inequality
\begin{equation}
2\mathnormal{\Gamma}_T\ge\mathnormal{\Gamma}_L
\label{eqn:2GT>=GL}
\end{equation}
holds at any temperature $T$ and for any $\varepsilon/\hbar\mathnormal{\Omega}_0$, while the equality $2\mathnormal{\Gamma}_T=\mathnormal{\Gamma}_L$ holds irrespectively of the parameters in the weak-coupling regime~\cite{ref:SLAinSpinBoson}.
Actually, the inequality~(\ref{eqn:2GT>=GL}) is proved on the basis of the master equation~(\ref{eqn:MasterEq}) or the Bloch equation~(\ref{eqn:BlochEq}).
For that purpose, the following proposition is available:
\begin{lemma}
\label{Proposition2}
\upshape If $A\in M_{3,R}$ satisfies the condition (\ref{eqn:i}), then the necessary and sufficient condition for the set of inequalities
\begin{eqnarray}
\Re\lambda_1+\Re\lambda_2&\ge&\Re\lambda_3,\nonumber\\
\Re\lambda_2+\Re\lambda_3&\ge&\Re\lambda_1,\label{eqn:GT}\\
\Re\lambda_3+\Re\lambda_1&\ge&\Re\lambda_2\nonumber
\end{eqnarray}
to hold is
\begin{equation}
f(\tr A/2)\ge0.
\label{eqn:ii}
\end{equation}
\end{lemma}
(Its proof is also elementary and straightforward~\cite{ref:GenInequalities}.)
Since the matrix $A$ in the Bloch equation~(\ref{eqn:BlochEq}) satisfies the condition~(\ref{eqn:ii}) as well as~(\ref{eqn:i}), the relaxation constants satisfy the inequalities~(\ref{eqn:GT}) by Prop.~\ref{Proposition2}\@.
When two of the three eigenvalues of $A$ are complex and the rest is real, the inequalities~(\ref{eqn:GT}) are nothing but the inequality~(\ref{eqn:2GT>=GL}).
Furthermore, the above proposition has shown that when there exist three longitudinal-relaxation constants $\mathnormal{\Gamma}_L^{(i)}$ ($i=1,2,3$), they satisfy the inequalities~(\ref{eqn:GT}), i.e.,
\begin{eqnarray}
\label{eqn:GG}
\mathnormal{\Gamma}_L^{(1)}+\mathnormal{\Gamma}_L^{(2)}&\ge&\mathnormal{\Gamma}_L^{(3)},\nonumber\\
\mathnormal{\Gamma}_L^{(2)}+\mathnormal{\Gamma}_L^{(3)}&\ge&\mathnormal{\Gamma}_L^{(1)},\\
\mathnormal{\Gamma}_L^{(3)}+\mathnormal{\Gamma}_L^{(1)}&\ge&\mathnormal{\Gamma}_L^{(2)}.\nonumber
\end{eqnarray}

The inequalities~(\ref{eqn:2GT>=GL}) and~(\ref{eqn:GG}) are also discussed by Gorini~\textit{et~al.}~\cite{ref:GoriniKossakowskiSudarshan}:
They investigated the Markovian master equation of the general form, i.e., the master equation of the Lindblad form~\cite{ref:LindbladCommunMathPhys1976,ref:GoriniKossakowskiSudarshan}, for two-level systems,
\begin{subequations}
\begin{equation}
\frac{d}{dt}\rho(t)=(\hat{\mathcal H}+\hat{\mathcal V})\rho(t),
\end{equation}
\begin{equation}
\hat{\mathcal H}\rho=-\frac{i}{\hbar}[H,\rho],\quad
\hat{\mathcal V}\rho=\frac{1}{2}\sum_{i,j=1}^{3}c_{ij}
\{[\sigma_i,\rho\sigma_j]+[\sigma_i\rho,\sigma_j]\}
\end{equation}
\end{subequations}
($H$ is an Hermitian operator in 2-dim.~Hilbert space and $\{c_{ij}\}$ is a positive semi-definite c-number matrix), and showed that the inequalities~(\ref{eqn:2GT>=GL}) and~(\ref{eqn:GG}) hold \textit{if the superoperators $\hat{\mathcal H}$ and $\hat{\mathcal V}$ commute}:
\begin{equation}
[\hat{\mathcal H},\hat{\mathcal V}]=0.
\label{eqn:Condi}
\end{equation}
In our master equation~(\ref{eqn:MasterEq}), which is also of the Lindblad form, they do not necessarily commute.
(Consider, for example, the case of $T>0$ and $\varepsilon\neq0$.)
This means that the inequalities~(\ref{eqn:2GT>=GL}) and~(\ref{eqn:GG}) hold more generally~\cite{note:IneqInGeneral}.
Notice that Eq.~(\ref{eqn:2GT>=GL}) is a well-known inequality in actual experiments~\cite{ref:MNRAbragam}.

It is true, of course, that the model Hamiltonian~(\ref{eqn:RapidDecayHamiltonian}) with~(\ref{eqn:SpinBosonModel}) is to be refined for a description of real experiments, but we expect the interesting and nontrivial features shown here to be universal ones which reflect some important characteristics related to nonlinear and/or nonequilibrium systems.

\begin{acknowledgments}
The authors acknowledge useful and helpful discussions with Profs.~Ohba, Nakazato, Tasaki, and Asahi.
One of the authors (K.~I.) is grateful to Luigi Accardi for kind hospitality.
He is supported by an overseas research fellowship of Japan Science and Technology Corporation.
\end{acknowledgments}



\begin{thebibliography}{19}
\bibitem{note:EquilibriumState}
If $\Re\lambda_i>0$, the solution of the Bloch equation~(\ref{eqn:BlochEq}) decays to zero, i.e., $\bm{D}(\tau)\to0$ as $\tau\to\infty$, and the state of the spin system, $\rho_S(\tau)$, approaches the state $\rho_\mathrm{eq}\propto1$, which is the thermal state at infinitely high temperature.
This is because the temperature of the boson system, $T$, is infinitely high for the spin system $\lambda^2H_S$ when $\lambda\to0$~\cite{ref:SLAForRapidDecay}.
\end{thebibliography}
\end{document}